\documentclass[10pt, letterpaper, onecolumn]{emulateapj}

\def\Msun{M_\odot}
\def\microas{\mu{\rm as}}

\begin{document}

\title {PHASES High Precision Differential Astrometry of $\delta$ Equulei}

\author{Matthew W.~Muterspaugh\altaffilmark{1}, Benjamin F.~Lane\altaffilmark{1}, Maciej Konacki\altaffilmark{2}, 
Bernard F.~Burke\altaffilmark{1}, M.~M.~Colavita\altaffilmark{3}, S.~R.~Kulkarni\altaffilmark{4}, M.~Shao\altaffilmark{3}}
\altaffiltext{1}{MIT Kavli Institute for Astrophysics and Space Research, MIT Department of Physics, 70 Vassar Street, Cambridge, MA 02139}
\altaffiltext{2}{Department of Geological and Planetary Sciences, California Institute of Technology, MS 150-21, Pasadena, CA 91125}
\altaffiltext{3}{Jet Propulsion Laboratory, California Institute of Technology, 4800 Oak Grove Dr., Pasadena, CA 91109}
\altaffiltext{4}{Division of Physics, Mathematics and Astronomy, 105-24, California Institute of Technology, Pasadena, CA 91125}

\email{matthew1@mit.edu, blane@mit.edu, maciej@gps.caltech.edu}

\begin{abstract}
$\delta$ Equulei is among the most well-studied nearby binary star systems.  
Results of its observation have been applied to a wide range of fundamental 
studies of binary systems and stellar astrophysics.  It is widely used to 
calibrate and constrain theoretical models of the physics of stars.  
We report 27 high precision differential astrometry measurements of 
$\delta$ Equulei from the Palomar High-precision Astrometric Search 
for Exoplanet Systems (PHASES).  The median size of the minor axes of the uncertainty ellipses for 
these measurements is 26 micro-arcseconds ($\microas$).  
These data are combined with previously 
published radial velocity data and other previously published differential 
astrometry measurements using other techniques
to produce a combined model for the system orbit.  
The distance to the system is determined to within a twentieth of a parsec and the 
component masses are determined at the level of a percent.  The constraints 
on masses and distance are limited by the precisions of the radial velocity data; 
we outline plans improve this deficiency and discuss the outlook for further study of this 
binary.
\end{abstract}


\section{Introduction}

The study of $\delta$ Equulei 
as a binary star has lasted nearly 200 years.  
In the early 1800s, William Herschel's (mistaken) 
listing of it as a wide binary (with what Friedrich Struve 
later proved to be an unrelated background star) 
brought it to the attention of many astronomers.  
While making follow-up observations to support 
his father's claim that the proposed pair 
were only an optical double, Otto Wilhelm von Struve 
in 1852 \nocite{o_struve1859} found that while the separation of 
the optical double continued to grow (to 32''), 
the point-spread-function of $\delta$ Equulei itself appeared elongated.  
He concluded that $\delta$ Equulei itself is a much more compact binary.  
(It is perhaps interesting to note that the Struve 
family's study of the system continued through 1955, 
when Otto Struve and K.~L.~Franklin \nocite{o_struve_1955} 
included the system in a spectroscopic study.)

$\delta$ Equulei (7 Equulei, HR 8123, HIP 104858, HD 202275, ADS 14773) 
is among the most well-studied nearby binary star systems.  
It is particularly useful to studies of binary systems and stellar 
properties as it is close ($d=18.4$ pc), bright ($V = 4.49$, $K = 3.27$), and 
can be studied both visually (semi-major axis roughly 
a quarter an arcsecond) and spectroscopically 
(spectral classes F7V+F7V, $K \approx 12.5~{\rm km\,s^{-1}}$) 
in a reasonable amount of time ($P \approx 5.7$ years); \cite{Mazeh1992} 
found only 23 binaries within 22 pc with periods less than 
3000 days and spectra that were nearly solar 
(spanning types F7-G9, classes IV-V, V, and VI).  
As such, it is regularly included in statistical surveys of 
binary systems \citep[see, for example, ][]{Heacox1998, Hale1994} and 
fundamental stellar properties such as the mass-luminosity relationship, 
calibrating photometric parallax scales, 
tabulating the H-R diagram of the solar neighborhood, 
and constraining models of stellar atmospheres 
\citep[see, for example, ][]{Lastennet2002, Eggen1998, Castelli1997, Smalley1995, Boehm1989, Habets1981, Popper1980}.  
These applications depend upon accurate knowledge of the components' 
physical properties and the system's parallax.

Previously, the visual orbit models 
(and thus evaluation of the total system mass and orbital parallax) 
of $\delta$ Equulei have been limited by 
differential astrometry with relative precisions on order of a few 
percent.  The recently developed method by two of us (see \cite{LaneMute2004a}) 
for ground-based differential astrometry at the 10 $\microas$ level for 
sub-arcsecond (``speckle'') binaries has been used to study $\delta$ Equulei over 
the 2003-2004 observing seasons.  These measurements represent an improvement 
in precision of over two orders of magnitude over previous work on this system.  
We present these new data, an updated three-dimensional model of the system, and 
the physical properties of the component stars.  
The observations were taken as part of the Palomar High-precision Astrometric 
Search for Exoplanet Systems (PHASES).  The primary goal of PHASES is to 
search for planets in sub-arcsecond binaries, with the data also being 
useful for studies of binaries themselves.  

PHASES data is collected at the Palomar Testbed Interferometer (PTI).
PTI is located on Palomar
Mountain near San Diego, CA \citep{col99}. 
It was developed by 
the Jet Propulsion Laboratory, California Institute of Technology for 
NASA, as a testbed for interferometric techniques applicable to the 
Keck Interferometer and other missions such as the Space 
Interferometry Mission, SIM.  
It operates in the J ($1.2 \mu{\rm 
m}$), H ($1.6 \mu{\rm m}$), and K ($2.2 \mu{\rm m}$) bands, and combines 
starlight from two out of three available 40-cm apertures. The 
apertures form a triangle with one 110 and two 87 meter baselines.

\section{Observations and Data Processing}

\subsection{PHASES Observations}

$\delta$ Equulei was observed with PTI on 27 nights in 2003-2004 
using the observing mode described in \cite{LaneMute2004a}.  We briefly 
review this method for phase-referenced differential astrometry of 
subarcsecond binaries here.

In an optical interferometer light is collected at two or more
apertures and brought to a central location where the beams are
combined and a fringe pattern produced.  For a broadband source of
central wavelength $\lambda$ the fringe pattern is limited in extent and
appears only when the optical paths through the arms of the
interferometer are equalized to within a coherence length ($\Lambda =
\lambda^2/\Delta\lambda$). For a two-aperture interferometer,
neglecting dispersion, the intensity measured at one of the combined
beams is given by
\begin{equation}\label{double_fringe_delEqu}
I(x) = I_0 \left ( 1 + V \frac{\sin\left(\pi x/ \Lambda\right)}
{\pi x/ \Lambda} \sin \left(2\pi x/\lambda \right ) \right )
\end{equation}
\noindent where x is the differential amount of path between arms of the 
interferometer, $V$ is the fringe contrast or ``visibility,'' which
can be related to the morphology of the source, and $\Delta\lambda$ is
the optical bandwidth of the interferometer assuming a flat optical
bandpass (for PTI $\Delta\lambda = 0.4 \mu$m). 

The location of the resulting interference fringes are 
related to the position of the target star and the observing geometry
via
\begin{equation}\label{delayEquation_delEqu}
d = \overrightarrow{B} \cdot \overrightarrow{S} + \delta_a\left(\overrightarrow{S}, t\right) + c 
\end{equation}
\noindent where d is the optical path-length one must introduce
between the two arms of the interferometer to find fringes. This
quantity is often called the ``delay.'' $\overrightarrow{B}$ is the baseline, 
the vector connecting the two apertures. $\overrightarrow{S}$ is the unit vector
in the source direction, and $c$ is a constant additional scalar delay
introduced by the instrument.  
The term $\delta_a\left(\overrightarrow{S}, t\right)$ 
is related to the differential amount of path introduced by the atmosphere 
over each telescope due to variations in refractive index.  
For a 100-m baseline interferometer an astrometric precision of 10 $\mu$as
corresponds to knowing $d$ to 5 nm, a difficult but not impossible 
proposition for all terms except that related to the atmospheric delay.  
Atmospheric turbulence, which changes over
distances of tens of centimeters, angles on order tens of arcseconds, 
and on subsecond timescales, forces
one to use very short exposures (to maintain fringe contrast) and
hence limits the sensitivity of the instrument. It also severely limits
the astrometric accuracy of a simple interferometer, at least over 
large sky-angles.

However, in narrow-angle astrometry we are concerned with a close pair
of stars, and we are attempting a differential astrometric
measurement, i.e.~we are interested in knowing the angle between the
two stars ($\overrightarrow{\Delta s} = \overrightarrow{s_2} - \overrightarrow{s_1} $).  
The atmospheric turbulence is correlated over
small angles.  If the measurements of the two stars are simultaneous, or nearly 
so, the atmospheric terms cancel.
Hence it is still possible to obtain high precision
``narrow-angle'' astrometry.

To correct for time-dependent fluctuations in the atmospheric turbulence, 
observations consisted of operating PTI in a phase-referenced observing 
mode.  After movable mirrors in the beam-combining lab apply delay 
compensation to the light collected from two 40 cm apertures,
the light from each aperture is split using 30/70 
beamsplitters.  Seventy percent of the light is sent to the 
phase-tracking ``primary'' interferometric beam combiner which 
measures the time-dependent phase of one star's interferogram (fringes) 
caused by the atmospheric turbulence, and used in a feed-forward 
loop to control the optical delay lines.

The other $30\%$ of the light is diverted to the ``secondary'' 
interferometric beam combiner.  In this system we have an additional 
delay line with a travel of only $\approx 500$ microns.  We use this 
to introduce delay with a sawtooth waveform with frequency on order 
a Hertz.  This allows us to sample the interferograms of all stars 
in the one arcsecond detector field whose projected separations 
are within the scan range.  Laser metrology is used along 
all starlight paths between the 30/70 split and the point of 
interferometric combination to monitor internal 
vibrations differential to the phase-referencing and scanning beam combiners.  
For $\delta$ Equulei, the typical scanning rate in 2003 was one scan per second and 
four intensity measurements per ten milliseconds; these values 
were doubled in 2004.  The typical scan amplitude was 100 microns.  An average of 1700 scans
were collected each night the star was observed over a time span of 30
to 90 minutes.

\subsection{PHASES Data Reduction}

The quoted formal uncertainties in the PHASES data are derived using 
the standard PHASES data reduction algorithm, which we review here.  
First, detector calibrations (gain, bias, and background) are applied to 
the intensity measurements.  Next, we construct a
grid in differential right ascension and declination over which to search 
(in ICRS 2000.0 coordinates).  
For each point in the search grid we calculate the expected
differential delay based on the interferometer location, baseline
geometry, and time of observation for each scan.  These conversions were simplified 
using the routines from  the Naval Observatory Vector Astrometry Subroutines 
C Language Version 2.0 (NOVAS-C; see \cite{novas}).  A model of a 
double-fringe packet is then calculated and compared to the observed
scan to derive a $\chi^2$ value as a merit of goodness-of-fit; this is repeated for each scan,
co-adding all of the $\chi^2$ values associated with that point in the
search grid. The final $\chi^2$ surface as a function of differential
right ascension and declination is thus derived. The best-fit astrometric
position is found at the minimum $\chi^2$ position, with uncertainties
defined by the appropriate $\chi^2$ contour---which depends on the
number of degrees of freedom in the problem and the value of the
$\chi^2$-minimum.

One potential complication with fitting a fringe to the data is that
there are many local minima spaced at multiples of the operating
wavelength. If one were to fit a fringe model to each scan separately
and average (or fit an astrometric model to) the resulting delays, one
would be severely limited by this fringe ambiguity (for a 110-m
baseline interferometer operating at $2.2 \mu$m, the resulting
positional ambiguity is $\sim 4.1$ milliarcseconds). However, by
using the $\chi^2$-surface approach, and co-adding the probabilities
associated with all possible delays for each scan, the ambiguity
disappears. This is due to two things, the first being that co-adding
simply improves the signal-to-noise ratio. Second, since the
observations usually last for an hour or even longer, the associated
baseline change due to Earth rotation also has the effect of ``smearing''
out all but the true global minimum. The final $\chi^2$-surface does
have dips separated by $\sim 4.1$ milliarcseconds from the true 
location, but any data sets for which these show up at the $4\sigma$ level 
are rejected.  The final astrometry measurement and related uncertainties 
are derived by fitting only the $4\sigma$ region of the surface.

The PHASES data reduction algorithm naturally accounts for contributions 
from photon and read-noise.  Unmonitored phase noise shows up by increasing 
the minimum value of $\chi^2$ surface.  Comparison of this value with 
that expected from the number of degrees of freedom allows us to co-add the 
phase noise to the fit.

This method has been rigorously tested on both synthetic and real data.  
Data sets are divided into equal sized subsets which are analyzed separately.  
A Kolmogorov-Smirnov test shows the formal uncertainties from the PHASES 
data reduction pipeline to be consistent with the scatter between subsets.  
After an astrometric solution has been determined, one can revisit the 
individual scans and determine best-fit delay separations on a scan-by-scan 
basis (the fringe ambiguity now being removed).
The differential delay residuals show normal (Gaussian )distribution, and Allan variances 
of delay residuals agree with the performance levels of the formal uncertainties 
and show the data to be uncorrelated.  We conclude that the PHASES data reduction 
pipeline produces measurement uncertainties that are consistent on intranight 
timescales.  Additionally, because the stars in $\delta$ Equulei are nearly identical (and in particular 
have equal temperatures to within the uncertainties of the best measurements), 
potential systematics such as differential dispersion are negligible.

The differential astrometry measurements are listed in 
Table \ref{phasesDeltaEquData}, in the ICRS 2000.0 reference frame.  
A Keplerian fit to the PTI data using the formal uncertainties 
found the minimized value of 
reduced $\chi_r^2=14.46$, implying either that the uncertainty 
estimates are too low by a factor of 3.8 or the (single) Keplerian 
model is not appropriate for this system.  Several possible sources of 
excess astrometric scatter have been evaluated.  
The PTI baseline vectors are determined by observing pointlike stars across the sky 
and inverting eq.~\ref{delayEquation_delEqu}.  The baseline 
solutions are stable at the level of a millimeter over year-long timescales, a factor of 
ten better than is required for 10 microarcsecond astrometry on subarcsecond binaries.  
To test the effect of the interferogram model template on the astrometry, the data were 
refit several times using several models for the instrumental bandpass; the 
differential astrometry solutions varied by less than a tenth of the formal uncertainties.
The maximum effect of starspots is evaluated as approximately 8 $\microas$ of 
scatter per 10 millimagnitudes of photometric variability, a level not observed 
in $\delta$ Equulei (Hipparcos photometry shows a scatter of only 4 
millimagnitudes \citep{hipPhotometry}).
The delay lines at PTI are in air rather than vacuum, introducing longitudinal 
dispersion to the system and color-dependent variations to the points of 
zero optical delay.  Because the components of $\delta$ Equulei are 
equal temperature, this effect cancels in a differential measurement.  
Because dispersion can couple to the astrometric measurement in more complicated 
ways (for example, if the optical alignment varies with either sky position or 
internal delay and the point spread function has color dependencies)
and some PHASES targets do not have equal temperature components, 
we are implementing a dispersion compensator for operation beginning 
in the 2005 observing season.  A more complete analysis of PHASES measurement 
uncertainties will be addressed in a future paper.

The uncertainty values 
listed in Table \ref{phasesDeltaEquData} have been increased by 
a factor of 3.8 over the formal uncertainties; these increased values 
are used in fits presented in this paper, in order that this data set 
can be combined with others.  At this time we do not find that 
more complicated models (such as adding additional unseen 
system components) produce better fits to the PTI data.
The rescaled (raw) median minor- and major-axis uncertainties are 26 (6.8) and 
465 (122) $\microas$.  The rescaled (raw) mean minor- and major-axis uncertainties 
are 35 (9.2) and 1116 (294) $\microas$.

\begin{table}
\begin{center}
Table \ref{phasesDeltaEquData}\\
PHASES data for $\delta$ Equulei\\
\begin{tabular}{llllllllll}
\tableline
\tableline
JD-2400000.5 & $\delta$RA    & $\delta$Dec  & $\sigma_{\rm min}$ & $\sigma_{\rm maj}$ & $\phi_{\rm e}$ &  $\sigma_{\rm RA}$ & $\sigma_{\rm Dec}$ & $\frac{\sigma_{\rm RA, Dec}^2}{\sigma_{\rm RA}\sigma_{\rm Dec}}$ & N \\
             &  (mas) & (mas) & ($\microas$)                  & ($\microas$)                  & (deg)     & ($\microas$)                  & ($\microas$)                  & & \\
\tableline
52896.18089 & 43.1331 & 28.3054 & 27.9 & 279.0 & 150.13 & 242.3 & 141.1 & -0.97371 & 854 \\
52897.15549 & 42.9731 & 29.2553 & 24.0 & 378.4 & 146.47 & 315.7 & 209.9 & -0.99054 & 1698 \\
52915.20224 & 46.9142 & 41.1237 & 34.8 & 465.2 & 163.69 & 446.6 & 134.8 & -0.96308 & 885 \\
52917.18203 & 46.8911 & 42.6169 & 15.0 & 273.3 & 160.60 & 257.8 & 91.8 & -0.98499 & 2197 \\
52918.17180 & 46.9129 & 43.3924 & 26.1 & 305.1 & 158.50 & 284.1 & 114.4 & -0.96945 & 1181 \\
52920.19763 & 46.8789 & 44.7316 & 37.2 & 1537.1 & 166.02 & 1491.6 & 373.0 & -0.99472 & 970 \\
52929.15827 & 49.5698 & 50.3713 & 18.0 & 333.9 & 162.42 & 318.4 & 102.3 & -0.98287 & 1846 \\
52930.16162 & 49.3723 & 51.1589 & 53.3 & 1344.3 & 163.85 & 1291.3 & 377.5 & -0.98912 & 584 \\
52951.10343 & 53.1325 & 64.3276 & 119.7 & 3391.8 & 163.63 & 3254.4 & 962.9 & -0.99157 & 232 \\
53172.44519 & 30.9090 & 105.4060 & 26.9 & 849.2 & 152.99 & 756.7 & 386.4 & -0.99695 & 1142 \\
53173.40171 & 30.6927 & 104.8840 & 34.4 & 629.2 & 17.94 & 598.7 & 196.6 & 0.98292 & 1322 \\
53181.39890 & 28.0318 & 101.4538 & 36.2 & 1586.5 & 149.30 & 1364.2 & 810.6 & -0.99865 & 679 \\
53182.39947 & 27.4793 & 101.1192 & 29.3 & 1339.1 & 149.87 & 1158.3 & 672.6 & -0.99873 & 1391 \\
53186.39977 & 25.5104 & 99.5860 & 69.3 & 2784.6 & 151.54 & 2448.4 & 1328.2 & -0.99824 & 250 \\
53187.39616 & 26.2104 & 98.5359 & 22.7 & 816.0 & 151.72 & 718.7 & 387.1 & -0.99778 & 2087 \\
53197.39430 & 21.9940 & 93.8953 & 15.7 & 126.8 & 155.88 & 115.9 & 53.8 & -0.94784 & 2668 \\
53198.36980 & 21.8513 & 93.2833 & 14.9 & 136.6 & 152.64 & 121.5 & 64.2 & -0.96511 & 3203 \\
53199.39127 & 22.2861 & 92.4201 & 66.5 & 2970.9 & 156.61 & 2727.0 & 1180.8 & -0.99812 & 515 \\
53208.31795 & 17.7859 & 87.9990 & 19.7 & 197.4 & 20.76 & 184.7 & 72.3 & 0.95684 & 3244 \\
53214.33956 & 15.9122 & 84.5147 & 22.5 & 210.8 & 154.02 & 189.8 & 94.5 & -0.96443 & 2792 \\
53215.33920 & 15.4419 & 83.9798 & 13.6 & 132.7 & 153.90 & 119.3 & 59.6 & -0.96725 & 3947 \\
53221.26266 & 14.0158 & 80.9103 & 12.6 & 240.5 & 15.60 & 231.7 & 65.8 & 0.98013 & 5719 \\
53222.34182 & 12.6377 & 79.9100 & 16.3 & 784.0 & 159.11 & 732.5 & 279.9 & -0.99806 & 2302 \\
53229.30103 & 10.0096 & 75.6266 & 16.6 & 407.2 & 154.97 & 369.0 & 172.9 & -0.99436 & 2316 \\
53236.23726 & 8.2328 & 70.6564 & 97.6 & 3640.5 & 147.94 & 3085.8 & 1934.1 & -0.99822 & 66 \\
53249.24027 & 1.7950 & 62.9850 & 14.9 & 192.5 & 153.47 & 172.4 & 87.0 & -0.98165 & 2234 \\
53251.19930 & 2.4844 & 60.8650 & 54.3 & 4775.1 & 148.21 & 4058.8 & 2516.1 & -0.99968 & 794 \\
\tableline
\end{tabular}
\caption[PHASES data for $\delta$ Equulei]
{ \label{phasesDeltaEquData}
All quantities are in the ICRS 2000.0 reference frame.  
The uncertainty values presented in these data have all been scaled 
by a factor of 3.8 over the formal (internal) uncertainties within 
each given night.  Column 6, $\phi_{\rm e}$, is the angle between the major axis of the 
uncertainty ellipse and the right ascension axis, measured from increasing differential 
right ascension through increasing differential declination (the position angle of the 
uncertainty ellipse's orientation is $90-\phi_{\rm e}$).  
The last column is the number of scans taken during a given night.  The quadrant 
was chosen such that the larger fringe contrast is designated the primary 
(contrast is a combination of source luminosity and interferometric visibility).
}
\end{center}
\end{table}

\subsection{Previous Differential Astrometry Measurements}

We have collected previously published differential astrometry 
measurements made with other methods.  Most of these measurements were 
tabulated by \cite{hart04} in the Fourth Catalog of Interferometric 
Measurements of Binary Stars, though several additional measurements 
(particularly those made by micrometer measures) had to be researched 
from the original sources.  In two cases we found discrepancies between the uncertainties 
listed in the Fourth Catalog and the original sources (the 1977.8811 point 
by \cite{BLM1980} and that from 1983.9305 by \cite{Bnu1984}); in each case we 
used the uncertainties listed in the original work.  Several data points listed 
without uncertainty estimates in the Fourth Catalog were found to have 
uncertainty estimates listed in the original works, in which case those values 
were used.

Most of the previous differential astrometry measurements were 
published without any associated uncertainties.  To allow these to be 
used in combined fits with other data sets, we determined average 
uncertainties as follows.  We separated 
the measurements into subgroups by observational method and 
analyzed each set individually; 
the first group included eyepiece and micrometer observations, 
and the second contained interferometric observations, including speckle, 
phase-grating, aperture masking, and adaptive optics.  The uncertainties were 
first estimated to be 10 milliarcseconds in separation and 
1 degree in position angle.  A Keplerian model was fit to the data, 
and residuals in separation and position angle treated individually 
to update the estimates and outliers removed.  This procedure 
was iterated until uncertainties 
were found consistent with the scatter.  The 66 visual data points used have 
average uncertainties of 37.2 milliarcseconds in separation and 
3.53 degrees.  The 58 interferometric data points used have average uncertainties 
of 5.92 milliarcseconds and 1.59 degrees.

We fit a Keplerian model to the data points for which uncertainty 
estimates were available to determine whether these were systematically 
too large or too small, and to find outliers.  Because there were only four 
visual/micrometer measurements with published uncertainties, we 
did not treat these as a separate group.  There were 42 interferometric 
measurements with published uncertainty estimates.  We found the 
uncertainty estimates to be systematically too small; this factor was 
larger in position angle than in separation.  Upon iteration, it was 
found that the separation uncertainties for these 46 data points 
needed to be increase by a factor of 1.71 and the position 
angle uncertainties by 2.38.

We list these previously published measurements in Tables 
\ref{prevWithUncertDeltaEquData}, \ref{visualWithoutUncertDeltaEquData}, 
and \ref{intWithoutUncertDeltaEquData} (complete 
versions available in the online version of this paper).

\nocite{WRH1941a} 
\nocite{Worley1957_AJ_62_153} 
\nocite{BLM1978} 
\nocite{McA1978b} 
\nocite{McA1982b} 
\nocite{BLM1980} 
\nocite{McA1980b} 
\nocite{Bnu1980b} 
\nocite{Tok1980} 
\nocite{Tok1982b} 
\nocite{Bag1984b} 
\nocite{Tok1983} 
\nocite{Bag1994} 
\nocite{Ism1992} 
\nocite{Ari1997} 
\nocite{Docobo1998_AAS_139_117} 
\nocite{Hor2001b} 
\nocite{Hrt2000a} 
\nocite{Hor1999} 
\nocite{Bag2002} 
\nocite{hor02} 

\nocite{Fin1960b} 
\nocite{vanDenBos1962_AJ67_141} 
\nocite{o_struve1859}
\nocite{B__1935a}
\nocite{WRH1941b}
\nocite{Couteau1962_JO_45_225} 
\nocite{Danjon1952_JdO_35_85}
\nocite{vanDenBos1963_AJ68_57} 
\nocite{Jef9999}
\nocite{Fin1963a} 
\nocite{Jef1945}
\nocite{Fin1963a} 
\nocite{WRH1950a}
\nocite{Fin1964a} 
\nocite{Muller1950_XXX}
\nocite{Muller1966_JO_49_335} 
\nocite{WRH1951}
\nocite{Fin1966a} 
\nocite{Fin1951b}
\nocite{Muller1966_JO_49_335} 
\nocite{WRH1952}
\nocite{Heintz1967_JO_50_343} 
\nocite{WRH1952}
\nocite{Couteau1968_JO_51_337} 
\nocite{Couteau1970_AAS_3_51} 
\nocite{Fin1953d}
\nocite{Holden1974_PASP_86_902} 
\nocite{Fin1954c}
\nocite{Holden1975_PASP_87_253} 
\nocite{Holden1976_PASP_88_325} 
\nocite{Holden1978_PASP_90_465} 
\nocite{WRH1955}
\nocite{Holden1979_PASP_91_479} 

\nocite{Hartkopf1996} 
\nocite{McA1983}
\nocite{Hrt1994} 
\nocite{Miu1993} 
\nocite{McA1984a}
\nocite{Hrt1997} 
\nocite{McA1987b}
\nocite{WSI1999a} 
\nocite{Bag1985}
\nocite{Hartkopf1996} 
\nocite{Tok1985}
\nocite{Bag1987}
\nocite{McA1987a}
\nocite{Bag1989a}
\nocite{McA1989}
\nocite{Hrt1997} 
\nocite{McA1990}
\nocite{WSI1999b} 
\nocite{Hrt1992b}
\nocite{Hrt1997} 
\nocite{Hrt1993}
\nocite{TtB2000}
\nocite{Iso1992}
\nocite{WSI1999b}

\begin{table}
\begin{center}
Table \ref{prevWithUncertDeltaEquData}\\
Previous Astrometry With Uncertainties\\
\begin{tabular}{llllll}
\tableline
\tableline
Year & $\rho$    & $\theta$  & $\sigma_{\rho}$ & $\sigma_{\theta}$ & Reference \\
     &  (mas)    & (deg) & (mas)           & (deg) & \\
\tableline
1934.578 & 161 & 191.9 & 8.55 & 4.52 & \cite{WRH1941a} \\
1934.826 & 100 & 170.2 & 17.10 & 18.09 & \cite{WRH1941a} \\
1955.55 & 380 & 203.4 & 17.10 & 3.09 & \cite{Worley1957_AJ_62_153} \\
1975.545 & 117 & 29.7 & 30.78 & 4.76 & \cite{BLM1978} \\
1976.4496 & 44 & 247 & 8.55 & 3.57 & \cite{McA1978b} \\
1976.6163 & 90 & 226.5 & 7.69 & 4.76 & \cite{McA1982b} \\
1976.6217 & 89 & 225.9 & 7.69 & 4.76 & \cite{McA1982b} \\
1977.6347 & 293 & 208.6 & 7.52 & 1.19 & \cite{McA1982b} \\
1977.8811 & 309 & 202.8 & 10.60 & 4.76 & \cite{BLM1980} \\
\tableline
\end{tabular}
\caption[Previous differential astrometry data with published uncertainties for $\delta$ Equulei]
{ \label{prevWithUncertDeltaEquData}
Previous differential astrometry data with published uncertainties for $\delta$ Equulei.  
$\rho$ uncertainties have been increased by a factor of 1.71 and those for $\theta$ by a factor of 2.38.  
In many cases $\theta$ has been changed by 180 degrees from the value appearing in the original works.  Complete version available at http://stuff.mit.edu/\~\,matthew1/deltaEquTables/.
}
\end{center}
\end{table}

\begin{table}
\begin{center}
Table \ref{visualWithoutUncertDeltaEquData}\\
Previous Visual Astrometry Without Uncertainties\\
\begin{tabular}{llll}
\tableline
\tableline
Year & $\rho$    & $\theta$  & Reference \\
     &  (mas)    & (deg)     &           \\
\tableline
1852.63 & 370 & 196.3 & \cite{o_struve1859} \\
1852.66 & 340 & 192.6 & \cite{o_struve1859} \\
1858.59 & 300 & 186.6 & \cite{o_struve1859} \\
1858.59 & 300 & 188.9 & \cite{o_struve1859} \\
1934.76 & 103 & 172 & \cite{B__1935a} \\
1937.46 & 260 & 212.6 & \cite{WRH1941b} \\
1938.88 & 296 & 198.7 & \cite{Danjon1952_JdO_35_85} \\
1939.513 & 262 & 200 & \cite{Jef1945} \\
1939.61 & 250 & 200.2 & \cite{Jef9999} \\
1939.690 & 250 & 199 & \cite{Jef1945} \\
\tableline
\end{tabular}
\caption[Previous visual differential astrometry data without published uncertainties for $\delta$ Equulei]
{ \label{visualWithoutUncertDeltaEquData}
Previous visual differential astrometry data without published uncertainties for $\delta$ Equulei.  
The uncertainties presented in this table were determined by the scatter in the data.
In many cases $\theta$ has been changed by 180 degrees from the value appearing in the original works.
For the values in this table, all uncertainties were taken to be $\sigma_{\rho} = 37.2$ mas and 
$\sigma_{\theta} = 3.53$ degrees.  Complete version available at http://stuff.mit.edu/\~\,matthew1/deltaEquTables/.
}
\end{center}
\end{table}

\begin{table}
\begin{center}
Table \ref{intWithoutUncertDeltaEquData}\\
Previous Interferometric Astrometry Without Uncertainties\\
\begin{tabular}{llll}
\tableline
\tableline
Year & $\rho$    & $\theta$  & Reference \\
     &  (mas)    & (deg)     &           \\
\tableline
1979.5296 & 261 & 197.9 & \cite{McA1982d} \\
1979.7700 & 227 & 195.4 & \cite{McA1982d} \\
1980.4772 & 78 & 176.1 & \cite{McA1983} \\
1980.4798 & 74 & 175.3 & \cite{McA1983} \\
1981.4628 & 131 & 22.6 & \cite{McA1984a} \\
1981.4710 & 132 & 21.2 & \cite{McA1984a} \\
1981.7032 & 97 & 14.1 & \cite{McA1984a} \\
1982.5059 & 136 & 217.3 & \cite{McA1987b} \\
1982.7600 & 189 & 211.6 & \cite{McA1987b} \\
1983.4232 & 292 & 206.7 & \cite{McA1987b} \\
\tableline
\end{tabular}
\caption[Previous interferometric differential astrometry data without published uncertainties for $\delta$ Equulei]
{ \label{intWithoutUncertDeltaEquData}
Previous interferometric differential astrometry data without published uncertainties for $\delta$ Equulei.
The uncertainties presented in this table were determined by the scatter in the data.
In many cases $\theta$ has been changed by 180 degrees from the value appearing in the original works.
For the values in this table, all uncertainties were taken to be $\sigma_{\rho} = 5.92$ mas and 
$\sigma_{\theta} = 1.59$ degrees.  Complete version available at http://stuff.mit.edu/\~\,matthew1/deltaEquTables/.
}
\end{center}
\end{table}

\subsection{Radial Velocity Data}

We have collected radial velocity data that has previously been published by 
\cite{dworetsky1971} and \cite{Popper1978} at Lick Observatory, 
\cite{Hans1979} at the Dominion Astrophysical Observatory (DAO), 
and \cite{duqMayor1988} from CORAVEL.  
We list these measurements in 
Tables \ref{lickDeltaEquData}, \ref{dominionDeltaEquData}, and \ref{coravelDeltaEquData} 
(full versions available in the online version of this paper).  
The Lick and DAO datasets were published without absolute uncertainty 
estimates but with relative weights assigned.  We fit each data set 
independently to a Keplerian model and use the scatter in the 
residuals to determine the absolute uncertainties, which we present 
in the tables.  A fit of the CORAVEL data to a Keplerian model showed 
excess scatter beyond the level of the formal uncertainties; we apply 
a scale factor of 1.527 to those uncertainties to allow these data to be 
combined with the other sets for simultaneous fits.

\begin{table}
\begin{center}
Table \ref{lickDeltaEquData}\\
Lick Observatory Radial Velocities\\
\begin{tabular}{lll}
\tableline
\tableline
JD-2400000.5 & RV 1                 & RV 2                 \\
             & (${\rm km\,s^{-1}}$) & (${\rm km\,s^{-1}}$) \\
\tableline
40410.5 & -6.8 & -24.7 \\
40428.3 & -5.2 & -24.9 \\
40430.3 & -4.9 & -26.0 \\
40459.3 & -3.4 & -27.3 \\
40487.2 & -1.6 & -28.8 \\
40494.3 & -1.7 & -29.4 \\
40516.2 & -0.8 & -31.4 \\
40544.1 & 1.3 & -31.9 \\
40691.5 & -1.7 & -29.7 \\
40731.5 & -4.5 & -26.9 \\
\tableline
\end{tabular}
\caption[Lick Observatory data for $\delta$ Equulei]
{ \label{lickDeltaEquData}
Lick Observatory data from \cite{dworetsky1971} and \cite{Popper1978} 
used for the combined orbital fit in this paper.  A model 
fit of this data set separately was used to determine the 
uncertainties used.  
Note that the original work switched the 
designations of the primary and secondary components; this has been 
corrected here for combination with other data sets.
For the values in this table, all uncertainties were taken to be 
0.35 ${\rm km\,s^{-1}}$, except for the point on ${\rm JD} - 2400000.5 = 40428.3$ 
for which the uncertainties were 0.70 ${\rm km\,s^{-1}}$.  Complete version available at http://stuff.mit.edu/\~\,matthew1/deltaEquTables/.
}
\end{center}
\end{table}

\begin{table}
\begin{center}
Table \ref{dominionDeltaEquData}\\
DAO Radial Velocities\\
\begin{tabular}{lllll}
\tableline
\tableline
Date    & RV 1                & $\sigma_1$          & RV 2                & $\sigma_2$              \\
(years) & (${\rm km\,s^{-1}}$) & (${\rm km\,s^{-1}}$) & (${\rm km\,s^{-1}}$) & (${\rm km\,s^{-1}}$) \\
\tableline
1966.770 & -21.96 & 0.81 & -9.70 & 0.81 \\
1966.902 & -22.05 & 0.81 & -9.73 & 0.81 \\
1967.461 & -21.56 & 0.81 & -10.17 & 0.81 \\
1967.593 & -20.99 & 0.81 & -9.87 & 0.81 \\
1969.415 & -9.62 & 0.81 & -22.18 & 0.81 \\
1969.437 & -8.73 & 0.81 & -23.01 & 0.81 \\
1969.456 & -8.24 & 0.41 & -23.69 & 0.41 \\
1969.502 & -6.96 & 0.41 & -24.48 & 0.41 \\
1969.565 & -5.86 & 0.41 & -26.12 & 0.41 \\
1969.568 & -5.38 & 0.41 & -25.46 & 0.41 \\
\tableline
\end{tabular}
\caption[Dominion Astrophysical Observatory data for $\delta$ Equulei]
{ \label{dominionDeltaEquData}
Dominion Astrophysical Observatory data from \cite{Hans1979}
used for the combined orbital fit in this paper.  A model 
fit of this data set separately was used to determine the 
uncertainties used.  Complete version available at http://stuff.mit.edu/\~\,matthew1/deltaEquTables/.
}
\end{center}
\end{table}

\begin{table}
\begin{center}
Table \ref{coravelDeltaEquData}\\
CORAVEL Radial Velocities\\
\begin{tabular}{lllll}
\tableline
\tableline
JD-2400000.5    & RV 1                & $\sigma_1$          & RV 2                & $\sigma_2$              \\
                & (${\rm km\,s^{-1}}$) & (${\rm km\,s^{-1}}$) & (${\rm km\,s^{-1}}$) & (${\rm km\,s^{-1}}$) \\
\tableline
43363.016 & -22.58 & 0.86 & -11.46 & 0.93 \\
43365.945 & -21.49 & 0.69 & -10.39 & 0.75 \\
43377.969 & -21.85 & 0.89 & -10.77 & 1.08 \\
43391.906 & -22.19 & 0.73 & -10.55 & 0.70 \\
43450.750 & -23.34 & 1.05 & -10.76 & 1.07 \\
43703.047 & -22.89 & 0.73 & -10.38 & 0.78 \\
43794.805 & -22.42 & 1.63 & -8.40 & 1.56 \\
43796.875 & -21.10 & 0.75 & -9.33 & 0.79 \\
43807.789 & -21.12 & 0.72 & -8.98 & 0.75 \\
44043.062 & -20.52 & 0.79 & -11.41 & 0.87 \\
\tableline
\end{tabular}
\caption[CORAVEL data for $\delta$ Equulei]
{ \label{coravelDeltaEquData}
CORAVEL data from \cite{duqMayor1988} used for the combined orbital fit in this paper.  
The uncertainties have been reweighted by a factor of 1.527 from the original work, 
in order that they might be combined with other data sets for a simultaneous fit.  
Note that the original work switched the 
designations of the primary and secondary components; this has been 
corrected here for combination with other data sets.  Complete version available at http://stuff.mit.edu/\~\,matthew1/deltaEquTables/.
}
\end{center}
\end{table}

\section{Orbital Models}

The first correct orbital solution for $\delta$ Equulei was that of 
\cite{luyten1934b}, and is consistent with the modern orbit.
\cite{vanDeKamp1945} measured the 
astrometry of the binary photocenter and derived its first photocentric orbit.
Their measurements yielded a measure of the mass ratio of 0.508:0.492 $\pm 0.016$ 
(\cite{vanDeKamp1954} later also derived individual 
masses of 1.96 and 1.89 $\Msun$ with the same method, values which are too 
large due to an underestimated parallax).  
The first spectroscopic orbit was by \cite{dworetsky1971}, providing a mass ratio of 
roughly 1.044.  Finally, a full three-dimensional model for the system was determined 
by \cite{Hans1979}.  Since that time, several more orbital solutions have 
been offered 
\citep[see, for example,][]{Starikova1981, duqMayor1988, Hartkopf1996, Soder1999, Pourbaix2000}.  

\begin{figure}[tbp]
\epsscale{0.6}
\plottwo{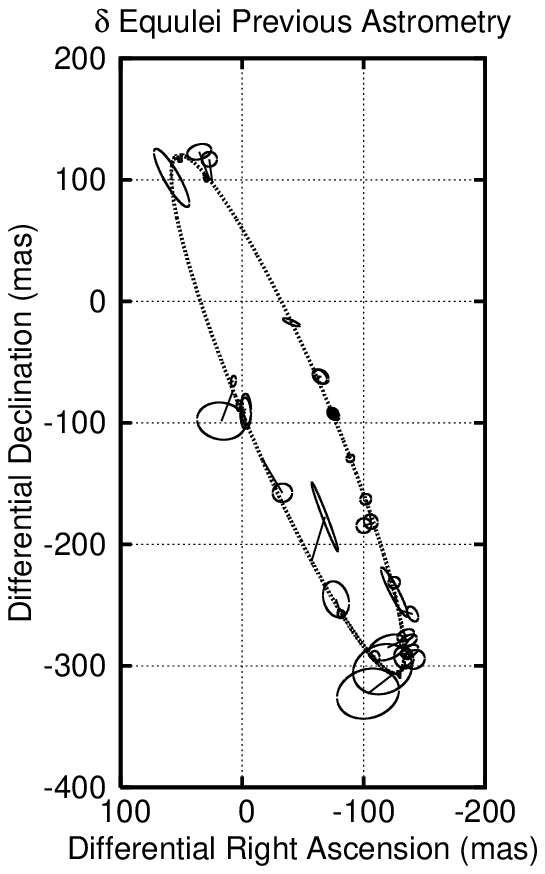}{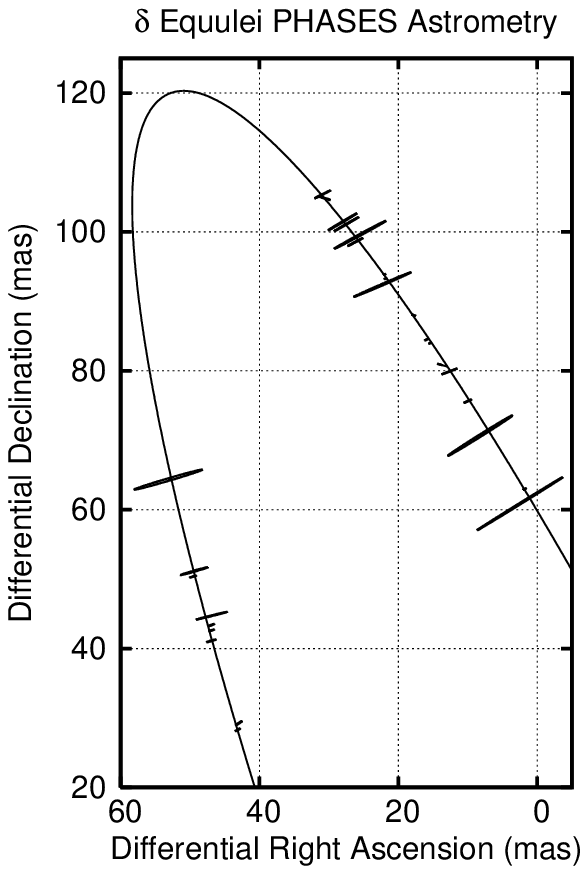}\\
\plotone{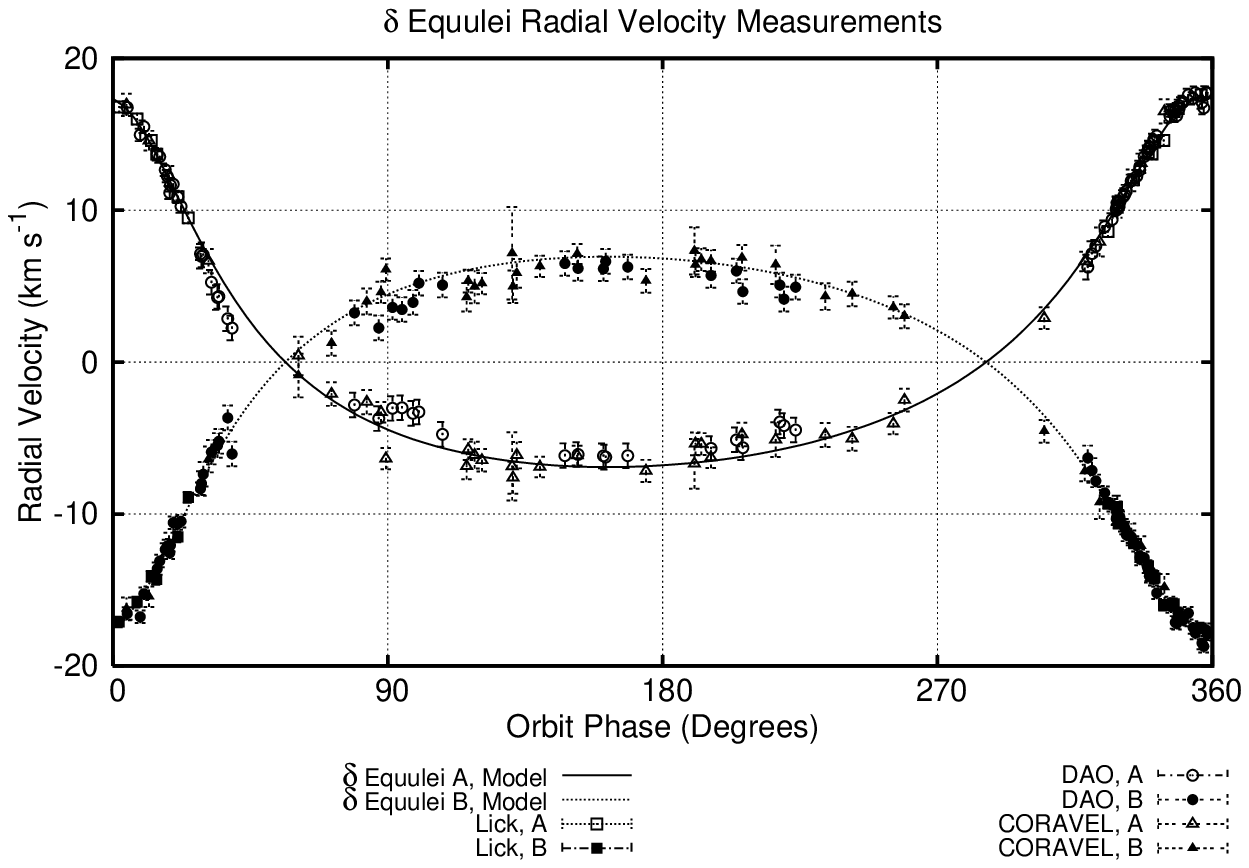}
\caption[The Orbit of $\delta$ Equulei] 
{ \label{delEquOrbit}
The orbit of $\delta$ Equulei.  
In the upper left plot, only interferometric differential astrometry measurements
for which the original works offered measurement uncertainty estimates 
have been included.  Note also that the dimensions of 
these uncertainty ellipses are those quoted by the original authors.
The dimensions of the uncertainty ellipses plotted for the PHASES 
measurements have been stretched by a factor of 3.8 as discussed in 
the text.  The high ellipticities of PHASES uncertainty 
ellipses are caused by use of a single baseline interferometer 
coupled with the limited range of hour angles over which $\delta$ 
Equulei could be observed (due to limited optical delay range at PTI).  
}
\end{figure}

We introduce three model parameters for the system velocity $V_o$, one corresponding to 
each observatory from which radial velocity data is obtained.  We do this to allow for 
instrumental variations; in particular, \cite{Hans1979} notes a potential zero-point 
discrepancy of 500 ${\rm m\,s^{-1}}$ in data sets.  Having fit each data set independently 
to correct uncertainty estimates, we combined all into a simultaneous fit to best 
determine system parameters.  The results are listed in Table \ref{deltaEquOrbitModels} and 
plotted in Figure \ref{delEquOrbit}.

Each fit was repeated several times varying the set of non-degenerate parameters used in order to 
obtain uncertainty estimates for a number of desired quantities.  The fit to 
radial velocity data alone was fit once using $a\sin i$ and $R = M_1/M_2$ as parameters, 
and again replacing these with $K_1$ and $K_2$, the velocity amplitudes.  Similarly, the 
combined fits were repeated replacing parameters $\{a, R\}$ with the 
sets $\{M=M_1+M_2, R\}$ and $\{M_1, M_2\}$.  Quantities in the combined fit that were 
derived from other parameters are listed separately at the end of Table \ref{deltaEquOrbitModels}.

Despite spanning less than a year (a sixth of the orbit), the PHASES data by themselves 
are able to constrain many orbital parameters better than previous observations.  We note 
in particular that the orbital angles are very well constrained.  However, the relatively 
short time coverage of the PHASES data presents strong degeneracies between the system 
period, eccentricity, and semi-major axis, which increases the fit uncertainties to 
levels much larger than one would expect given the precision of the astrometry.  
If, for example, one holds fixed the period and eccentricity at the fit values, the 
uncertainty in semi-major axis drops from 3200 to 92 $\microas$.  

It is noted that the PHASES only fit results in values of 
period, eccentricity, and semimajor axis that agree with previous fits 
at only the $3\sigma$ level.  At this time the causes of 
these discrepancies are not yet known.  As noted, these parameters are 
degenerate with each other in the PHASES only fit, thus it is not 
surprising that all are discrepant at the same level.
However, the data sets agree well 
enough to be simultaneously fit in a combined model 
(for which parameter degeneracies are less of a problem), and the resulting 
parameters do agree well with previous orbital models.  

We find that the addition of the previous differential astrometry to the combined model does 
little to improve the fit.  The combined fit is limited by the precision of the radial 
velocity observations.  Using the technique for obtaining high precision radial velocimetry 
on double line spectroscopic binaries using an iodine cell reported in \cite{Konacki04}, 
one of us (MK) is leading an effort to obtain such data to better constrain the orbital 
model, component masses, and system distance.  The PHASES program will also continue to observe 
$\delta$ Equulei so that the combination of high precision radial velocities and differential 
astrometry can be used for a comprehensive search for giant planets orbiting either star.  
We have conducted simulations which show collecting ten radial velocity 
measurements with 20 ${\rm m\,s^{-1}}$ precisions during the 2005 observing season will 
improve the constraints on the component masses by a factor of two.

\begin{table}
\begin{center}
{\tiny
Table \ref{deltaEquOrbitModels}\\
Orbit models for $\delta$ Equulei.\\
\begin{tabular}{l|llllll}
\tableline
\tableline
 & \cite{Soder1999} & \cite{Pourbaix2000} & RV & PHASES & PHASES & PHASES $+$ \\*
 &   &   &    &        & $+$ RV & Pre.~$+$ RV\\*
\tableline
$P$ (yr)                          & 5.713            & 5.703 $\pm 0.0070$  & 5.706 $\pm 0.003$    & 5.40 $\pm 0.11$      & 5.7059 $\pm 0.0003$     & 5.7058 $\pm 0.0003$     \\
$P$ (days)                        &                  &                     & 2084.08 $\pm 0.92$   & 1974 $\pm 39$        & 2084.07 $\pm 0.12$      & 2084.05 $\pm 0.11$      \\
$T_{0}$ (yr)                      & 1992.85          & 1981.47 $\pm 0.012$ & 2004.285 $\pm 0.015$ & 2004.299 $\pm 0.002$ & 2004.2954 $\pm 0.001$   & 2004.2950 $\pm 0.001$   \\
$T_{0}$ (MJD)                     &                  &                     & 53109.9  $\pm 5.5$   & 53114.53 $\pm 0.75$  & 53112.90 $\pm 0.45$     & 53112.76 $\pm 0.42$     \\
$e$                               & 0.44             & 0.440 $\pm 0.0046$  & 0.4519 $\pm 0.0029$  & 0.415 $\pm 0.0080$   & 0.437001 $\pm 0.000076$ & 0.436983 $\pm 0.000072$ \\
$a$ (mas)                         & 231              & 232 $\pm 1.8$       &                      & 222.8 $\pm 3.2$      & 231.85 $\pm 0.11$       & 231.88 $\pm 0.11$       \\
$V_{0, Lick}$ (${\rm km\,s^{-1}}$) &                 &                     & -15.398 $\pm 0.097$  &                      & -15.40 $\pm 0.10$       & -15.40 $\pm 0.11$       \\
$V_{0, DAO}$ (${\rm km\,s^{-1}}$)  &                 &                     & -15.876 $\pm 0.074$  &                      & -15.875 $\pm 0.078$     & -15.875 $\pm 0.081$     \\
$V_{0, C}$ (${\rm km\,s^{-1}}$)    &                 & -15.85 $\pm 0.074$  & -15.728 $\pm 0.095$  &                      & -15.73 $\pm 0.10$       & -15.73 $\pm 0.10$       \\
$M_1$ ($\Msun$)                   &                  & 1.19 $\pm 0.034$    &                      &                      & 1.192 $\pm 0.012$       & 1.193 $\pm 0.012$       \\
$M_2$ ($\Msun$)                   &                  & 1.12 $\pm 0.032$    &                      &                      & 1.187 $\pm 0.011$       & 1.188 $\pm 0.012$       \\
$M_1+M_2$ ($\Msun$)               & 2.35             &                     &                      &                      & 2.380 $\pm 0.018$       & 2.380 $\pm 0.019$       \\
$M_1/M_2$                         & 1.06 $\pm 0.03$  & 1.06 $\pm 0.018$    & 1.004 $\pm 0.011$    &                      & 1.004 $\pm 0.012$       & 1.004 $\pm 0.012$       \\
$i$ (deg)                         & 99               & 99.0 $\pm 0.43$     &                      & 99.520 $\pm 0.052$   & 99.394 $\pm 0.020$      & 99.396 $\pm 0.019$      \\
$\omega$ (deg)                    & 3                & 8.0 $\pm 1.0$       & 187.01 $\pm 0.91$    & 188.53 $\pm 0.25$    & 187.96 $\pm 0.12$       & 187.92 $\pm 0.11$       \\
$\Omega$ (deg)                    & 203              & 203.8 $\pm 0.29$    &                      & 203.292 $\pm 0.044$  & 203.301 $\pm 0.046$     & 203.312 $\pm 0.043$     \\
$\pi$ (mas)                       & 54.32 $\pm 0.90$ & 55.0 $\pm 0.67$     &                      &                      & 54.38 $\pm 0.14$        & 54.39 $\pm 0.15$        \\
$d$ (pc)                          &                  &                     &                      &                      & 18.388 $\pm 0.048$      & 18.386 $\pm 0.050$      \\
\tableline
$a$ (AU)                          &                  &                     &                      &                      & 4.263 (Derived)         & 4.263 (Derived)     \\
$a \sin i$ (AU)                   &                  &                     & 4.186 $\pm 0.011$    &                      & 4.206 (Derived)         & 4.206 (Derived)     \\
$K_1$ (${\rm km\,s^{-1}}$)        &                  &                     & 12.224 $\pm 0.074$   &                      & 12.181 (Derived)        & 12.182 (Derived)    \\
$K_2$ (${\rm km\,s^{-1}}$)        &                  &                     & 12.272 $\pm 0.074$   &                      & 12.229 (Derived)        & 12.230 (Derived)    \\
\tableline
\end{tabular}
\caption[Orbital models for $\delta$ Equulei]
{ \label{deltaEquOrbitModels}
Pre.:  Previous differential astrometry measurements, listed in Tables 
\ref{prevWithUncertDeltaEquData}, \ref{visualWithoutUncertDeltaEquData}, 
and \ref{intWithoutUncertDeltaEquData}.\\
In the combined PHASES $+$ previous $+$ RV fit, all parameter uncertainties 
have been increased by a factor of $\sqrt{\chi_r^2} = \sqrt{1.17}$ (though the 
$\chi_r^2$ of the combined fit is artificial due to rescaling the 
uncertainties of the individual data sets, this reflects the degree to which 
the data sets agree with eachother).
}
}
\end{center}
\end{table}

\newpage
\section{Parallax}

Early attempts to measure the parallax of $\delta$ Equulei 
suffered from systematic errors due to its binary nature until 
\citeauthor{luyten1934b} determined a model for the visual orbit.
In the same paper that \citeauthor{vanDeKamp1945} determined the 
first orbit of the $\delta$ Equulei photocenter, they determined 
a trigonometric parallax of $48 \pm 5$ milliarcseconds.  

The best current values for the trigonometric parallax of 
$\delta$ Equulei are given by \cite{Gatewood1994} as an average 
of ground based observations ($54.2 \pm 0.93$ mill-arcseconds) 
and from the Hipparcos mission (a binary-orbit corrected parallax of 
$54.32 \pm 0.90$ milliarcseconds is reported \citep{Soder1999}).
Our combined orbital solution using PHASES differential astrometry and 
previously published radial velocity measurements provides the best 
estimate of orbital parallax, at $ 54.39 \pm 0.15$ milliarcseconds, in good 
agreement with the trigonometric values and the previous best 
orbital parallax of $55 \pm 0.67$ milliarcseconds \citep{Pourbaix2000}.
The orbital parallax determination is limited by the precision of the radial velocity 
measurements; simulations show that the high precision radial velocity 
observations planned for the next observing season will improve the precision 
by a factor of two.

\section{System Age}

\begin{figure}[tbp]
\epsscale{0.5}
\plotone{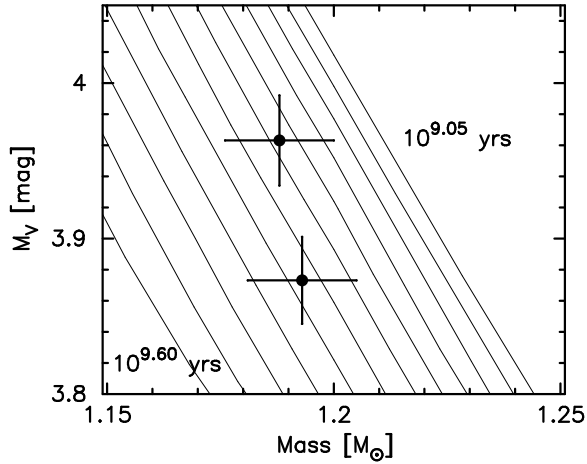}
\caption[Mass-Luminosity Isochrones for $\delta$ Equulei] 
{ \label{delEquIsochrones}
Isochrones for stars of near-solar metallicities as functions of 
stellar mass and absolute magnitude.  Isochrones shown are separated by 
$\log t = 0.05$.  Also plotted are the properties of 
the two components of $\delta$ Equulei.  The system age is 
$\log t  = 9.35^{+0.1}_{-0.15}$.
}
\end{figure}

The measured apparent $V$ magnitude for the system is reported 
as $4.487\pm 0.02$ by the Simbad astronomical database.  Combining this 
value with the measured $\Delta V$ between the stars of $0.09\pm0.04$ measured 
by \cite{TtB2000} using adaptive optics on the Mt.~Wilson 100'' telescope and 
the distance determined by our orbital model, we find the components have 
absolute magnitudes of $V_1 = 3.87\pm0.028$ and $V_2 = 3.96\pm0.029$.
These values are combined with the stellar evolution models of \cite{Girardi2000} 
to determine the system age.  The system's metallicities (in solar units) 
of $\left[ {\rm Z/H} \right] = -0.07$ \citep{Gray2003} 
and $\left[ {\rm Fe/H} \right] = -0.07$ \citep{Nordstrom2004} most closely match
\citeauthor{Girardi2000}'s isochrone for stars of solar metallicity.  
We find the system age is $\log t  = 9.35^{+0.1}_{-0.15}$ ($\approx 2.2\pm 0.6$ Gyr).  
The relevant isochrones are plotted in Figure \ref{delEquIsochrones}.

\section{$\delta$ Equulei and PHASES}

$\delta$ Equulei is a sample system discussed in the S-type 
(orbiting just one stellar component of a binary) planet stability 
studies of \cite{Rabl1988} and \cite{holman1999}.  The numerical simulations 
of \citeauthor{Rabl1988} determined that planets were stable around either star if 
their orbital semi-major axis were 0.68 AU ($P = 0.34$ year) 
or smaller; an additional semi-stable region existed out to 0.86 AU.  
The conclusion of \citeauthor{holman1999} was that the regions of stability were 
of size 0.67 AU ($P=0.43$ years; they assumed slightly different values for the 
component masses) around the primary and 0.66 AU around the secondary ($P=0.42$ years).  
From these studies, we assume a stable region of roughly $\frac{2}{3}$ AU around each 
star in which planets could be found.

\begin{figure}[tbp]
\epsscale{0.9}
\plottwo{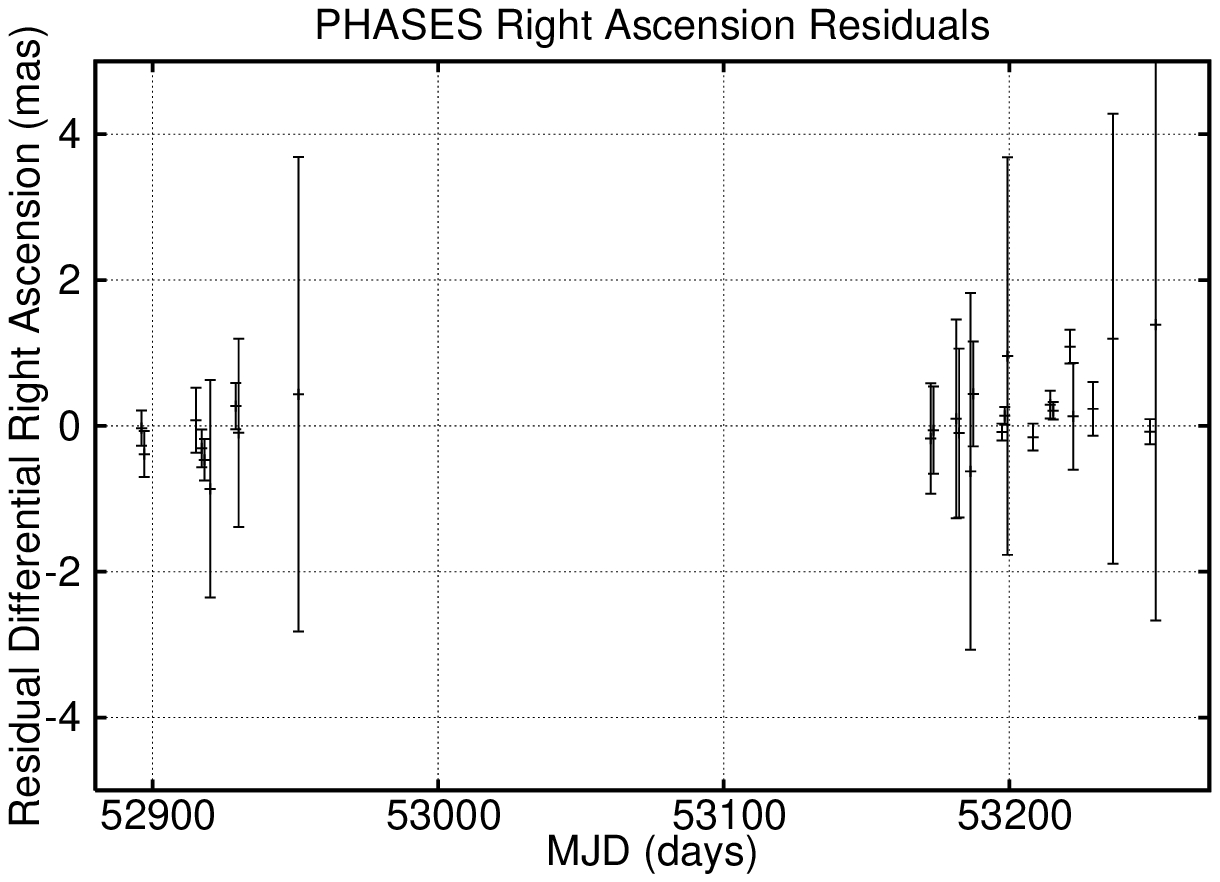}{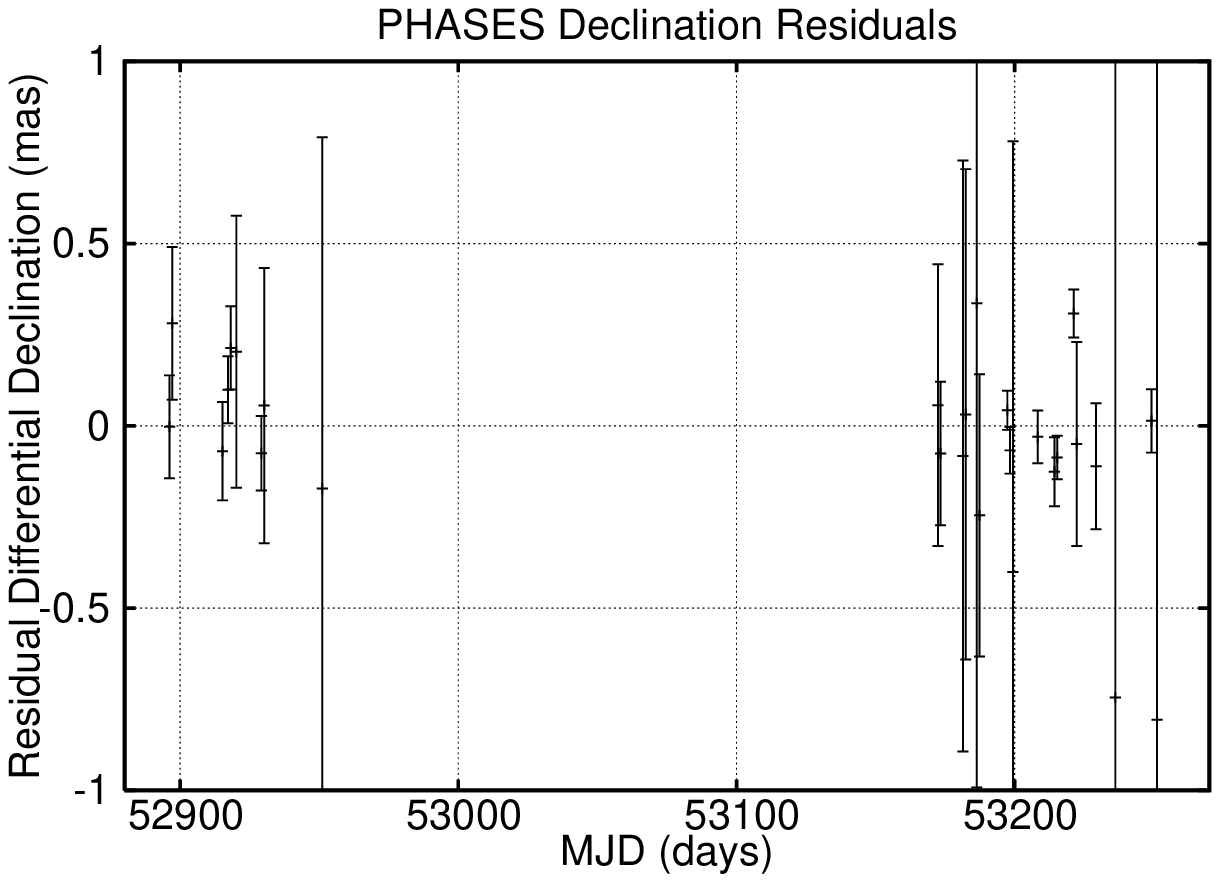}\\
\plottwo{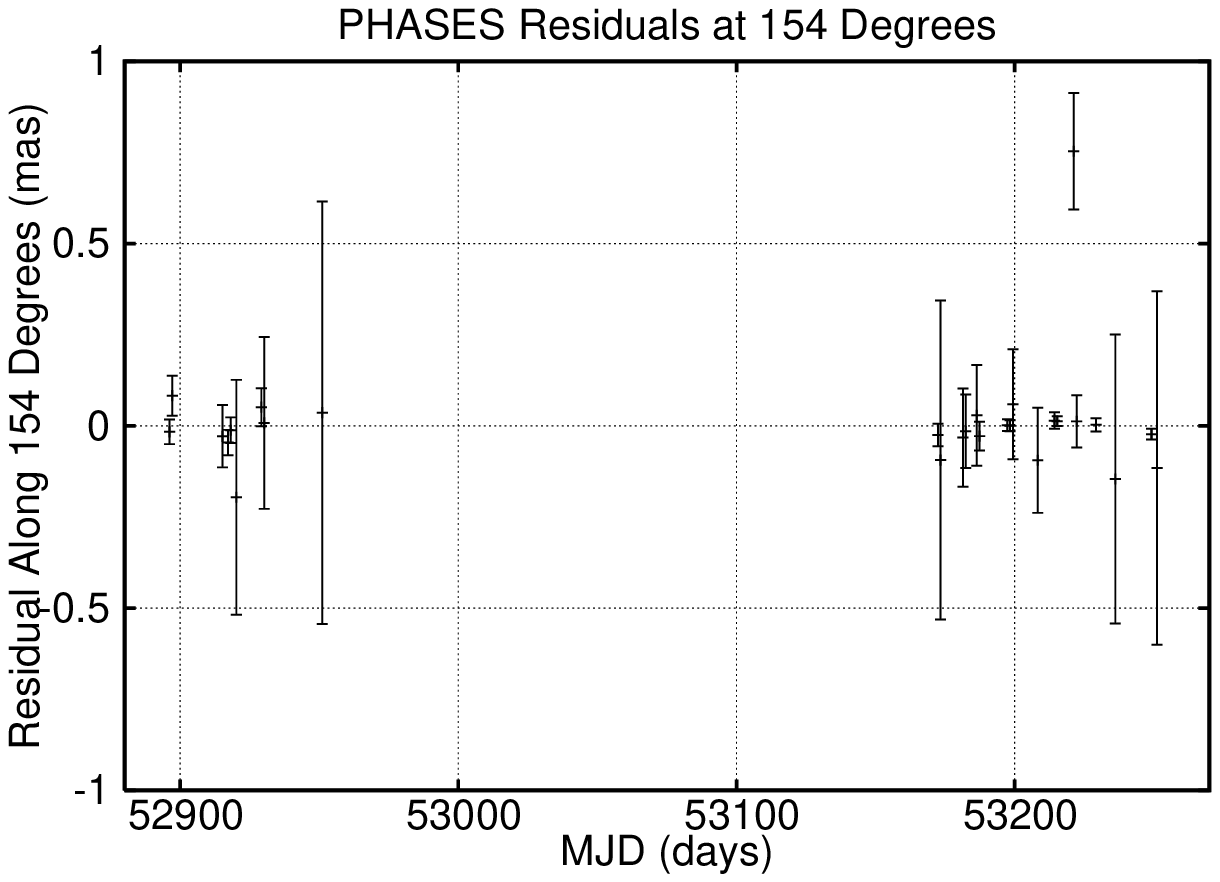}{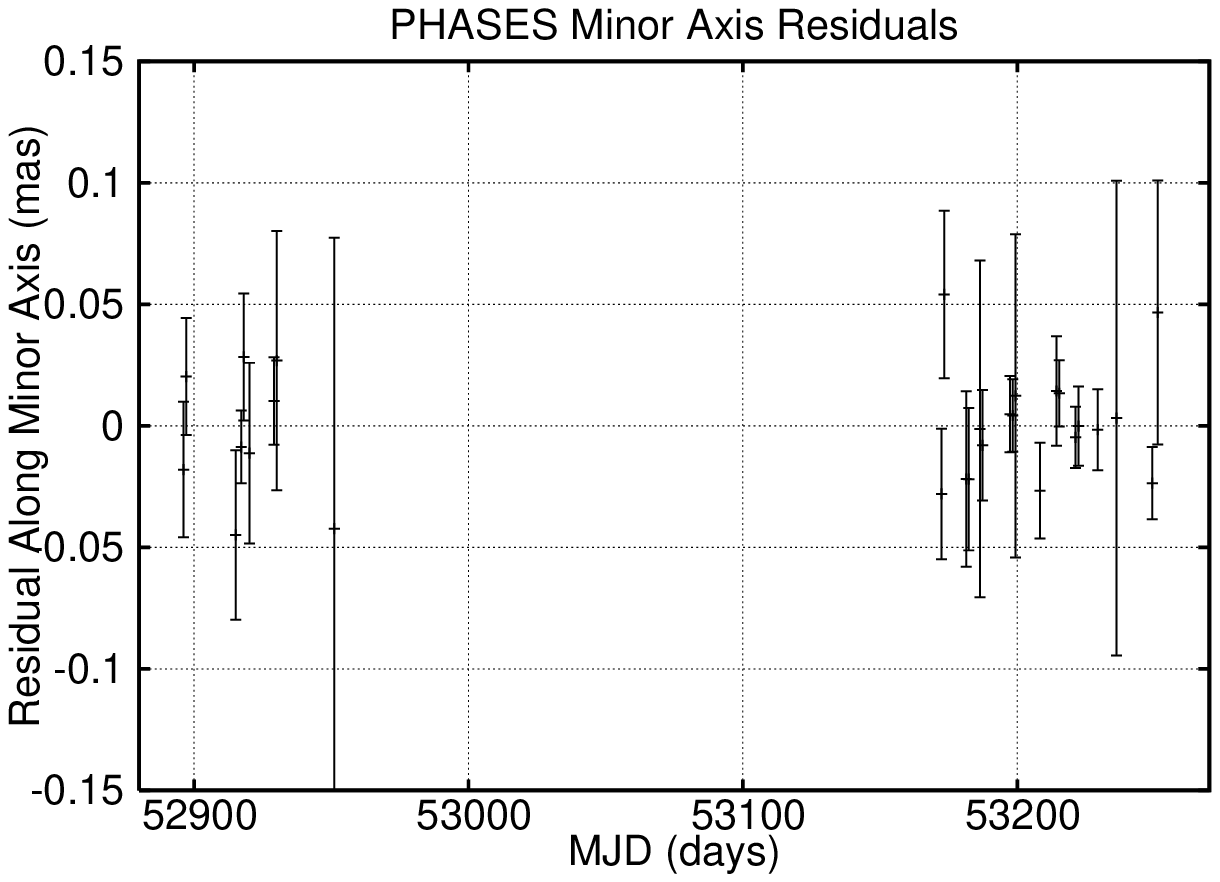}
\caption[Residuals for PHASES differential astrometry of $\delta$ Equulei] 
{ \label{delEquPhasesResiduals}
Residuals for PHASES differential astrometry of $\delta$ Equulei.
The error bars plotted have been stretched by a factor of 3.8 over the formal 
uncertainties as discussed in the text.  
The high ellipticity of the uncertainty ellipses 
causes neither the right ascension nor the declination uncertainties to be near the precision 
of the minor axis uncertainties, which have median uncertainty of 26 $\microas$.
Due to the roughly North-South alignment of the baseline used for 24 of the 27 measurements, 
our more sensitive axis was typically declination.
The bottom left plot shows the residuals along a direction that is 154 degrees from increasing differential 
right ascension through increasing differential declination 
(equivalent to position angle 296 degrees), which corresponds to 
the median direction of the minor axis of the PHASES uncertainty ellipses.  Because the orientation 
of the PHASES uncertainty ellipses varies from night to night, no single axis is ideal 
for exhibiting the PHASES precisions, but this median axis is best aligned to do so.
The bottom right plot 
shows residuals along the minor axis of each measurement's uncertainty ellipse.
}
\end{figure}

\begin{figure}[tbp]
\epsscale{0.6}
\plotone{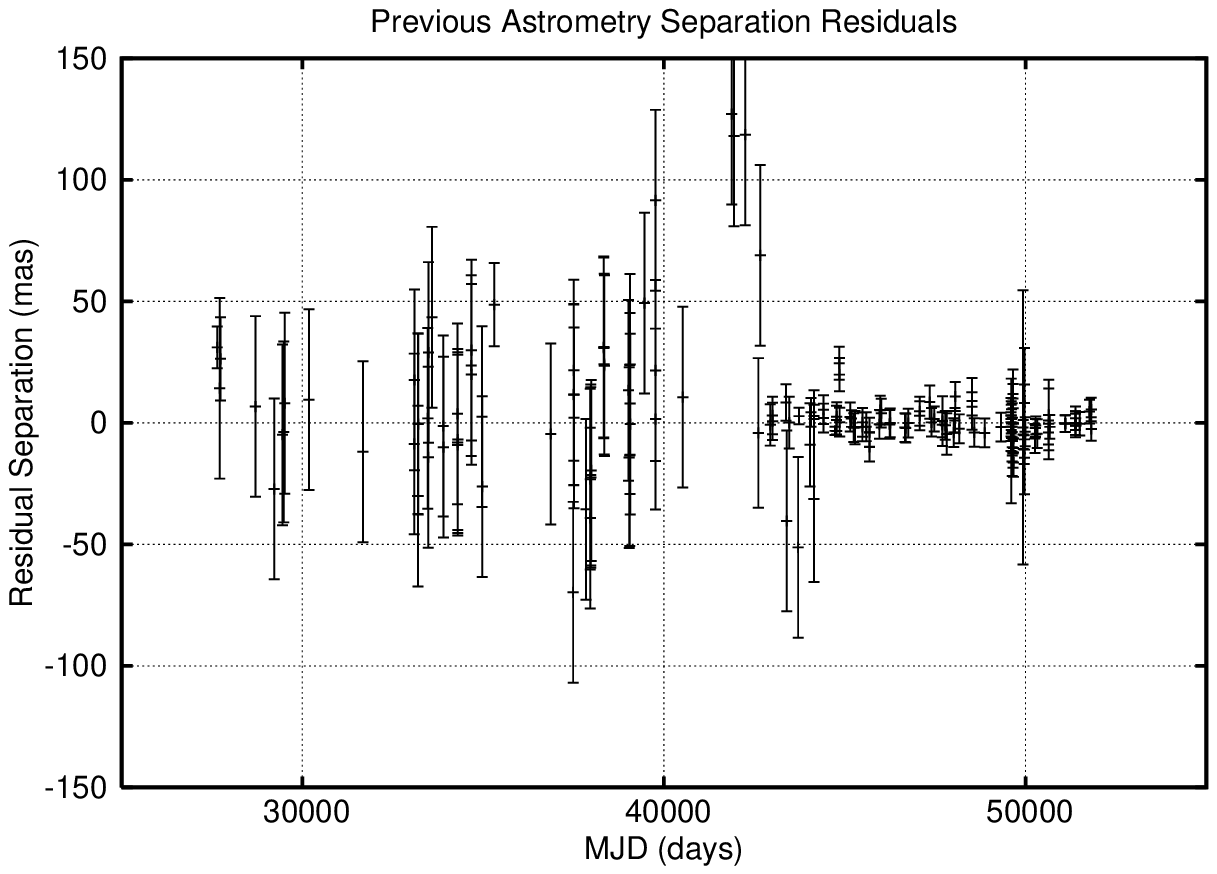}
\plotone{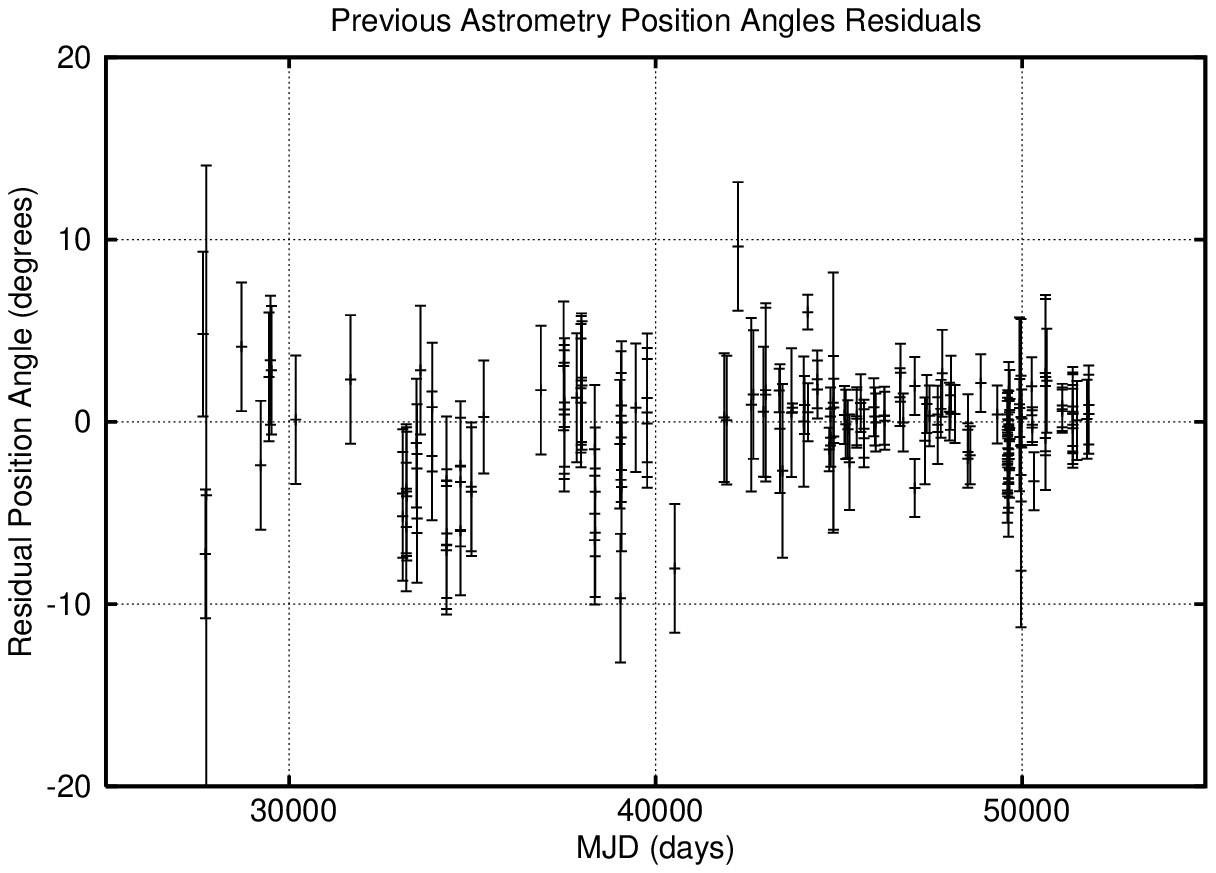}
\caption[Residuals for previous differential astrometry of $\delta$ Equulei] 
{ \label{delEquPreviousResiduals}
Residuals for previous differential astrometry of $\delta$ Equulei.
We have not included four points from the 1850's by Otto Wilhelm von Struve, though they also fit well.
}
\end{figure}

\begin{figure}[tbp]
\epsscale{0.6}
\plotone{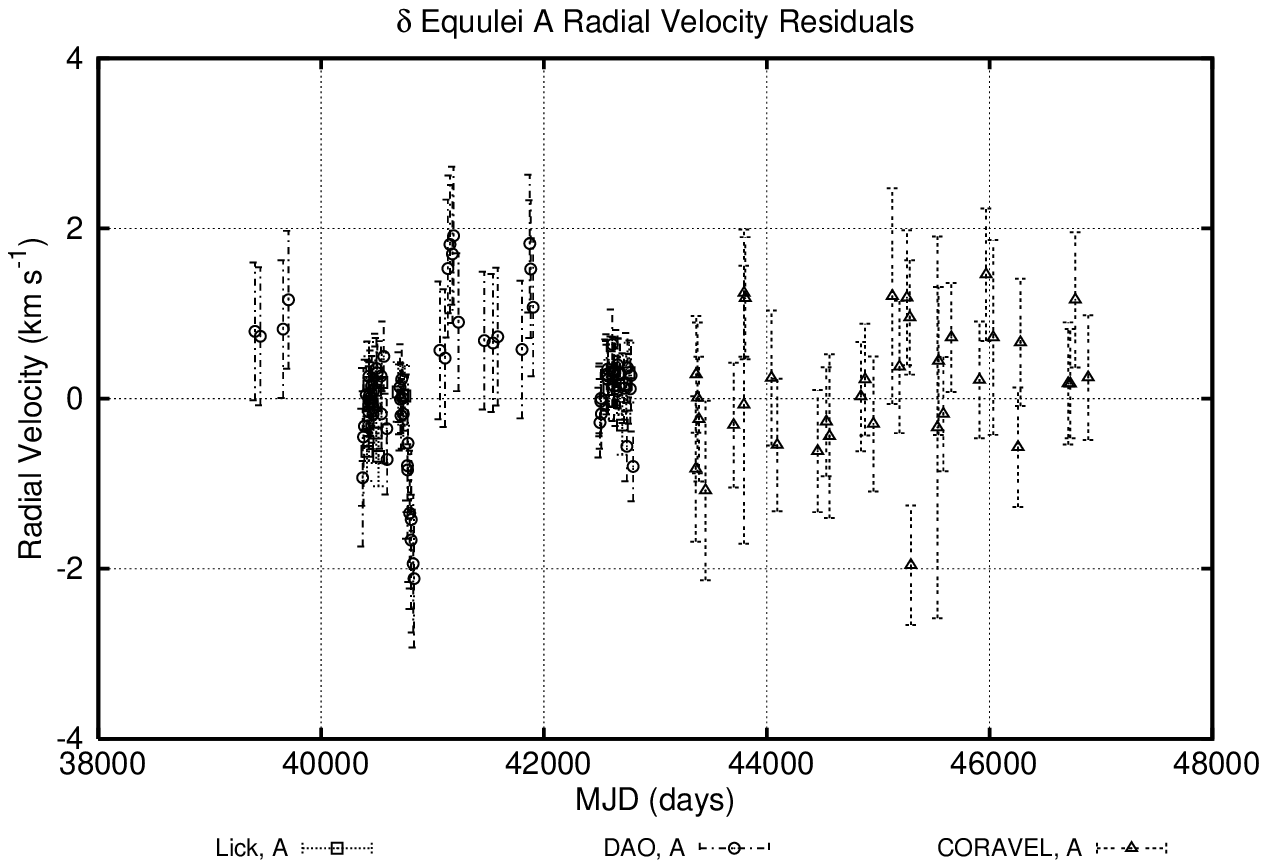}
\plotone{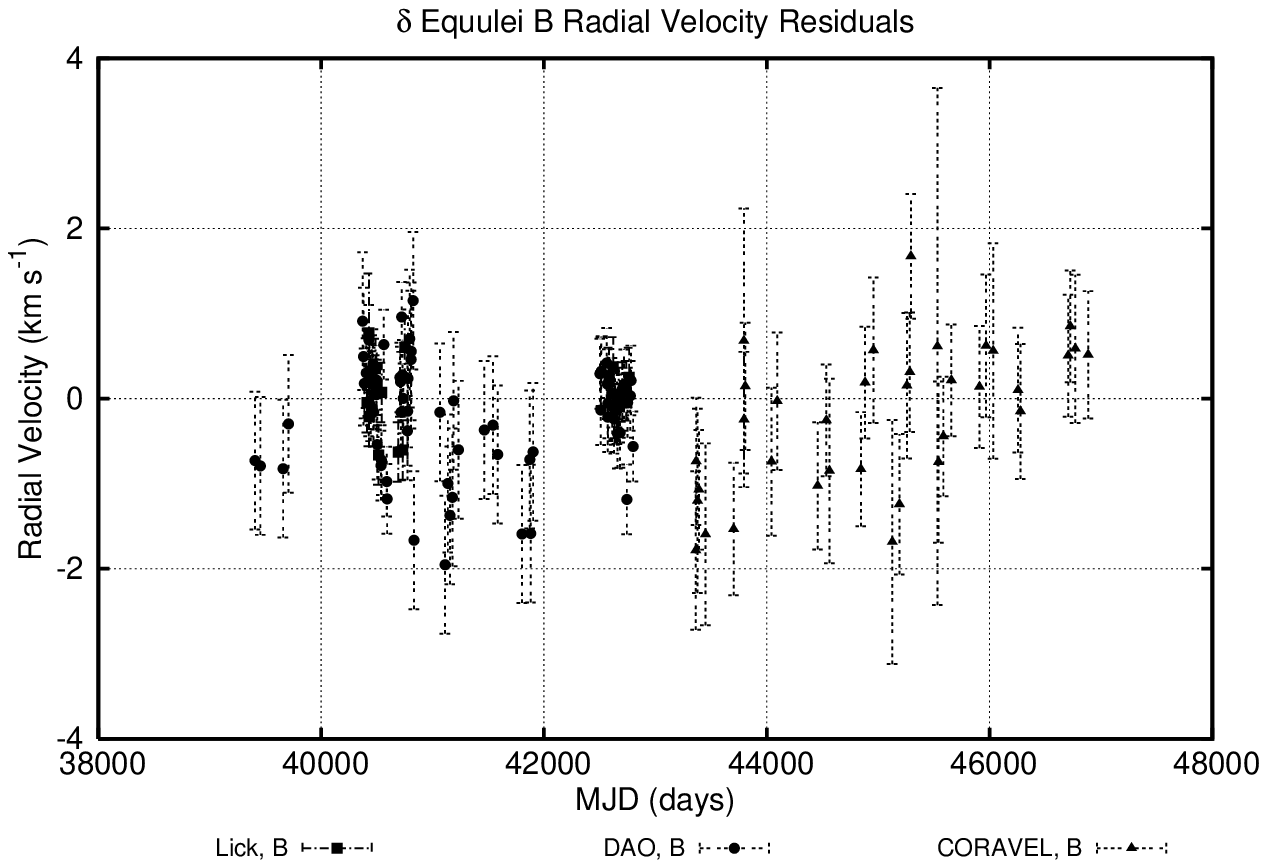}
\caption[Residuals for previous differential astrometry of $\delta$ Equulei] 
{ \label{delEquRVResiduals}
Residuals for radial velocimetry of $\delta$ Equulei, from three 
observatories.
}
\end{figure}

We do not find any obvious periodicities in our residuals, which 
are plotted in Figures \ref{delEquPhasesResiduals}, \ref{delEquPreviousResiduals}, 
and \ref{delEquRVResiduals}.  
Periodograms of the PHASES fit residuals show no clear peaks 
between one and 180 days.  We attempted to refit the PHASES data 
using a double-Keplerian model; we started each attempt with seed values 
for the wide Keplerian portion equal to the values found for the single 
Keplerian fit, and chose 3500 different starting values between 
one and 200 days for the period of the narrow portion (secondary Keplerian).  
The final value for the reduced $\chi_r^2$ was never found to fall below 
12.69, which is not significantly different from the value from the single 
Keplerian model of 14.46; because several seed periodicities produced 
$\chi_r^2$ near the 12.69 level, we conclude this slight dip is a result 
of random noise and the data sampling function.  We can safely conclude 
that there are no periodic signals in our residuals at the level of 
$100 \microas$, at least along the average minor axis of our uncertainty ellipses.  
At this conservative level we can conclude that there are not additional companions of mass 
\begin{equation}
M_p \ge 11.5 \left(\frac{P}{\rm month}\right)^{-\frac{2}{3}}~{\rm Jupiter~Masses}.
\end{equation}
The orbit of a third body could be hidden if it happens to be high inclination and 
coaligned with the major axis of our uncertainty ellipses.  A more thorough analysis 
of the fit residuals and better constraints on companion masses will be part of a 
future investigation.

\section{Conclusions}

The high precision differential astrometry measurements of the PHASES program 
are used to constrain the distance to $\delta$ Equulei more than four times more 
precisely than previous studies, despite covering only a sixth of the orbit.  
The orbital parallax agrees well with the trigonometric one determined by 
Hipparcos observations.  Whereas characterization of the system was 
previously limited by the precisions of differential astrometry 
measurements, it is now limited by the radial velocity observations.  
Continued monitoring of this nearby standard binary will be useful to 
search for additional system components as small as a Jupiter mass in 
dynamically stable orbits.

\acknowledgements 
We thank the support of the PTI collaboration, whose members have contributed 
designing an extremely reliable instrument for obtaining precision astrometric measurements.  
We acknowledge the extraordinary efforts of K. Rykoski, whose work in
operating and maintaining PTI is invaluable and goes far beyond the
call of duty.  Observations with PTI are made possible through the
efforts of the PTI Collaboration, which we acknowledge. Part of the
work described in this paper was performed at the Jet Propulsion
Laboratory under contract with the National Aeronautics and Space
Administration. Interferometer data was obtained at the Palomar
Observatory using the NASA Palomar Testbed Interferometer, supported
by NASA contracts to the Jet Propulsion Laboratory. This research has
made use of the Simbad database, operated at CDS, Strasbourg,
France. MWM acknowledges the support of the Michelson Graduate
Fellowship program. BFL acknowledges support from a Pappalardo
Fellowship in Physics.  
MK is supported by NASA through grant NNG04GM62G.
PHASES is funded in part by the California 
Institute of Technology Astronomy Department.

\bibliography{main}
\bibliographystyle{plainnat}

\end{document}